\documentclass{aa}

\usepackage{soul}
\usepackage{graphicx}
\usepackage{color}
\usepackage{txfonts}
\usepackage{natbib}
\usepackage[squaren,Gray]{SIunits}

\usepackage{bbold}
\usepackage{subcaption}
\usepackage{mwe}

\newcommand{\sink}{\mathrm{sink}}

\newcommand\nmhd{$\textsc{no\_outflows}$}
\newcommand\nmhds{$\textsc{no\_outflows}$ }

\newcommand\nmhdj{$\textsc{outflows}$}
\newcommand\nmhdjs{$\textsc{outflows}$ }

\begin{document}

   \title{Influence of protostellar outflows on star and protoplanetary disk formation in a massive star-forming clump}
   \author{U. Lebreuilly$^{1}$
   \and P. Hennebelle$^{1}$
    \and A. Maury$^{1}$
     \and M. González$^{2}$
\and A. Traficante$^{3}$
\and R. Klessen$^{4,5}$
 \and L. Testi$^{6,7}$
\and S. Molinari$^{3}$}
   \institute{$^{1}$Universit\'{e} Paris-Saclay, Universit\'{e} Paris Cité, CEA, CNRS, AIM, 91191, Gif-sur-Yvette, France\\
       $^{2}$Universit\'{e} Paris Cité, Universit\'{e} Paris-Saclay, CEA, CNRS, AIM, F-91191, Gif-sur-Yvette, France\\ 
         $^{3}$INAF - Istituto di Astrofisica e Planetologia Spaziali (INAF-IAPS), Via Fosso del Cavaliere 100, I-00133, Roma, Italy\\
   $^{4}$Universität Heidelberg, Zentrum für Astronomie, Institut für theoretische Astrophysik, Albert-Ueberle-Str. 2, D-69120 Heidelberg, Germany\\ 
      $^{5}$Universität Heidelberg, Interdisziplinäres Zentrum für Wissenschaftliches Rechnen, Im Neuenheimer Feld 205, D-69120 Heidelberg, Germany \\
       $^{6}$Dipartimento di Fisica e Astronomia ”Augusto Righi” Viale Berti Pichat 6/2, Bologna \\
    $^{7}$INAF, Osservatorio Astrofisico di Arcetri, Largo E. Fermi 5, I-50125, Firenze, Italy \\
\email{ugo.lebreuilly@cea.fr}    
}

   \date{}

  \abstract{ Due to the presence of magnetic fields, protostellar jets/outflows are a natural consequence of accretion onto protostars. They are expected to play an important role for star and protoplanetary disk formation.}{We aim to determine the influence of outflows on star and protoplanetary disk formation in star forming clumps.}{Using \texttt{RAMSES}, we perform the first  magnetohydrodynamics calculation of massive star-forming clumps with ambipolar diffusion, radiative transfer including the radiative feedback of protostars and protostellar outflows while systematically resolving the disk scales. We compare it to a model without outflows. }{We find that protostellar outflows have a significant impact on both star and disk formation. They provide significant additional kinetic energy to the clump, with typical velocities of a few $\sim 10 ~\kilo\meter~\second^{-1}$, impact the clump and  disk temperatures, reduce the accretion rate onto the protostars and enhance fragmentation in the filaments. We find that they promote a more numerous stellar population. They do not impact much the low mass end of the IMF, which is probably controlled by the mass of the first Larson core, however, that they have an influence on its peak and high-mass end. }{Protostellar outflows appear to have a significant influence on both star and disk formation and should therefore be included in realistic simulations of star-forming environments. }
      \authorrunning{Lebreuilly et al}
  \maketitle
  
\section{Introduction}

Fast ejection of matter is an expected consequence of magnetized flows through  magneto-centrifugal launching mechanisms \citep{1982MNRAS.199..883B}. These so-called jets/outflows are ubiquitous around protostars. Molecular outflows are observed with typical velocities of a few $10~\kilo\meter~\second^{-1}$, while strongly collimated jets can reach even higher $>100 ~\kilo\meter~\second^{-1}$ velocities  \citep[see the reviews by][]{2014prpl.conf..451F,2023ASPC..534..567P}. 

In numerical models, outflows are either launched self-consistently or through sub-grid modeling. The latter method is often used to investigate their large scale influence for isolated collapse \citep[e.g.,][]{2017ApJ...847..104O,2022MNRAS.510.2552R} as well as for star-forming clouds \citep[e.g.,][]{2009ApJ...695.1376C,2010ApJ...709...27W,2011ApJ...740..107C,2014ApJ...790..128F,2018MNRAS.475.1023M,2018MNRAS.473.4220L,2021MNRAS.502.3646G,2022A&A...663A...6V,2022MNRAS.512..216G,2023MNRAS.518.5190M} models. Generally, outflows have been shown to play an important role for star formation, reducing the star formation rate and perhaps even controlling the peak of the stellar initial mass function (IMF) by setting a mass scale, as proposed by several studies \citep{2018MNRAS.473.4220L,2021MNRAS.502.3646G,2023MNRAS.518.5190M}.

At small scales, protoplanetary disks are expected to form through angular momentum conservation. As the progenitors of planets, they are very important astrophysical objects. Over the past few years, significant efforts  to resolve disk populations in massive star-forming clumps have been done by the community. Such numerically costly calculations were first performed by \cite{2018MNRAS.475.5618B} including a treatment of the radiative transfer but without accounting for magnetic fields. \cite{2021MNRAS.508.5279E} later investigated the impact of the metallicity on disk formation with similar models. The role of magnetic fields on disk formation was investigated first by \cite{Kuffmeier2017} and \cite{Kuffmeier2019} for such simulations. They however did not systematically resolve the disk scales, but rather zoomed-in on a few specific disks. Disk populations were then explored while accounting for non-ideal MHD effects for low- \citep{2019MNRAS.489.1719W} and high- \citep{Lebreuilly2021} mass star-forming clumps. In \cite{Lebreuilly2021}, we have shown that magnetic fields play an important role in regulating the typical disk size in clumps as well as the number of stars in the cloud. In \cite{Hennebelleetal2022}, we have shown that they were also affecting the shape of the IMF, as strongly magnetized clouds would produce a stellar population with a more top heavy IMF that weakly magnetized ones. So far, none of these models accounted for protostellar outflows.

 In this paper, we present the first star-forming clump simulation that includes ambipolar diffusion, radiative transfer (including the feedback from the star internal and accretion luminosity) and a sub-grid modeling of protostellar outflows as implemented by \cite{2022A&A...663A...6V} while systematically resolving the disks up to $\sim 1~$au. We aim to study the impact of outflows on both star and disk formation and on the clump evolution. We thus compare this simulation with a model without outflows.

The plan of the article is as follows. In Sect.~\ref{sec:methods}, we present our methods, that are largely similar those of our previous works \citep{Lebreuilly2021}.  In Sect.~\ref{sec:results}, we discuss our results, describing the impact of protostellar outflows. In particular, we present their impact on the density and velocity structure of the cloud, on the population of protoplanetary disks as well as on star formation. Then, we present our conclusions in Sect.~\ref{sec:conclusions}.

\section{Methods}
\label{sec:methods}
\begin{figure}
  \centering
 \includegraphics[width=0.35
         \textwidth]{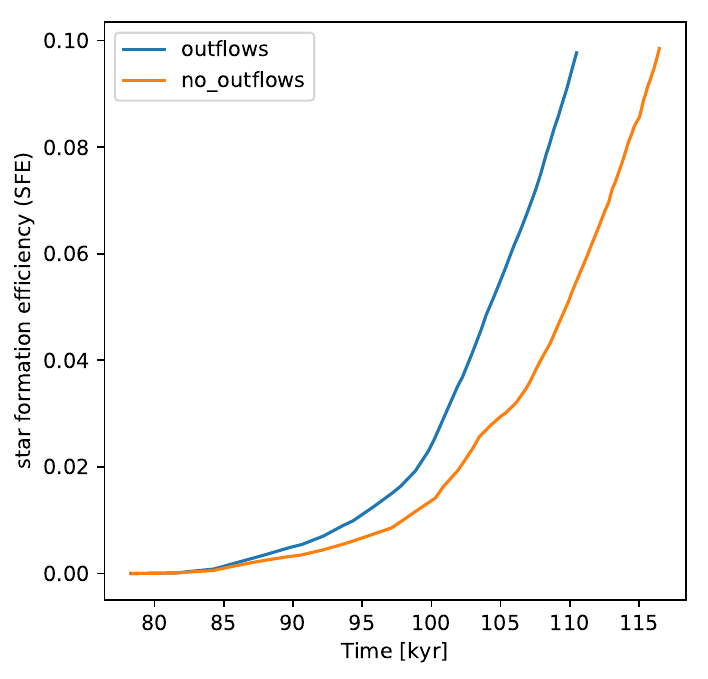}
    \caption{\label{fig:SFE} SFE as a function of time for both models. Outflows significantly slow down star formation near the cluster. }
\end{figure}

In this section, we describe the essential aspects of our methods. We point out that, outflows and resolution apart, the calculations presented here are the same as the $\textsc{nmhd}$ run from \cite{Lebreuilly2021}.

\subsection{Setup}

\label{sec:setup}

We computed our models using the \texttt{RAMSES} code \citep{RAMSES,2006A&A...457..371F} and its extension to radiation hydrodynamics \citep{2011A&A...529A..35C}, non-ideal MHD \citep{2012ApJS..201...24M} and sink particles \citep{2014MNRAS.445.4015B}. As in \cite{Lebreuilly2021}, we initially consider a $1000~ M_{\odot}$ clump with an uniform density. The corresponding  size $R_0$ of the clumps of temperature $T_0=10~$K are initialised according to the thermal-to-gravitational energy ratio $\alpha =0.08$ as
\begin{equation}
\alpha \equiv \frac{5}{2} \frac{R_0 k_{\rm{B}} T_{0}}{\mathcal{G} M_0\mu_{\rm{g}}m_{\rm{H}}},
\end{equation}
were we define the gravitational constant $\mathcal{G}$, the Boltzmann constant $k_{\rm{B}}$, the Hydrogen atom mass $m_{\rm{H}}$ and the mean molecular weight $\mu_{\rm{g}}=2.31$. This leads to an initial density of $\sim 3 \times 10^{-19} \gram ~\centi\meter^{-3}$. In addition, we initialise the velocity with supersonic turbulent fluctuations at Mach $\mathcal{M} = 7$ with a powerspectrum of $k^{-11/3}$.

Finally, we assume an initially uniform vertical magnetic field fixed according to the mass-to-flux over critical-mass-to-flux ratio $\mu=10$ as 
\begin{equation}
    \mu = \left(\frac{M_0}{\phi}\right)/\left(\frac{M}{\phi}\right)_c
\end{equation}
the critical mass-to-flux ratio being $\left(\frac{M}{\phi}\right)_c = \frac{0.53}{\pi}\sqrt{5/\mathcal{G}}$ \citep{1976ApJ...210..326M}. This corresponds to a magnetic field strength of $9.4 \times 10^{-5}~$G. We point out that we only consider the impact of ambipolar diffusion here, the resistivity is computed using the table from \cite{2016A&A...592A..18M}. The Ohmic dissipation is expected to play a role at densities higher than those we resolve here. In addition, the Hall effect could play a significant role but also unfortunately remains too challenging to account for in such simulations.

To accurately follow the gas dynamics up to the scales of protoplanetary disks we take advantage of the Adaptive Mesh Refinement (AMR) grid \citep{1984JCoPh..53..484B} of \texttt{RAMSES}. In this work, we refine the grid according to a modified Jeans length defined as 
 \begin{equation}
    \tilde{\lambda_{\mathrm{J}}}= \left\{\begin{array}{ll}
    \lambda_{\mathrm{J}}  & \mathrm{~if ~} n< 10^9 ~\centi\meter^{-3},\\
    \mathrm{min} (\lambda_{\mathrm{J}},\lambda_{\mathrm{J}} (T_{\mathrm{iso}}=300~ \kelvin)) & \mathrm{~otherwise ~}
    \end{array}
\right.
\end{equation}
This modification in the refinement criterion with respect from \cite{Lebreuilly2021} was done to ensure that the hot disk component is always refined up to the maximal resolution. Here, we impose to have 10 points per modified Jeans length up to a resolution of 1.2 au. This is enough to avoid artificial fragmentation of the cloud according to the criterion of \cite{1997ApJ...489L.179T} while being computationally affordable. We point out that our coarser cells are $\sim 2460~$au. 

\subsection{Implementation of the protostellar outflows}
\label{sec:outflows_methods}

We use sink particles to mimic protostars. We refer the reader to \cite{2014MNRAS.445.4015B} for more extensive details of their implementation. As in \cite{Lebreuilly2021}, we consider that a fraction of the accretion luminosity $f_{\mathrm{acc}}$ is radiated away by protostars under the form of an accretion luminosity that writes
\begin{equation}
    L_{\rm{acc}} = f_{\rm{acc}} \frac{\mathcal{G} M_{\sink} \dot{M}_{\sink}}{R_{\star}}.
\end{equation}

As in \cite{Lebreuilly2021}, we consider the case $f_{\mathrm{acc}}=0.1$. The stellar radius, $R_{\star}$, is computed using the evolutionary tracks of \cite{2013ApJ...772...61K}. As explained by \cite{2022A&A...658A..52C}, these pre-main sequence (PMS) tracks were computed using the STELLAR radiation code. In our case, we use a tabulated version of these tracks as a function of the mass accretion rate (their different models) onto the sink and the stellar mass. The initial object assumed by these models is $0.05 M_{\odot}$, has a radius of $0.56 R_{\odot}$ and a luminosity of $2.2 \times 10^{-2} L_{\odot}$.

In the case of our model with protostellar outflows, we have employed a sub-grid modelling  previously implemented in \texttt{RAMSES} by \cite{2022A&A...663A...6V}. We recall here the general principle of the methods. As in \cite{Lebreuilly2021}, the sinks of our model accrete, at each timestep, a fraction of the mass above the density threshold $n_{\mathrm{thre}}= 10^{13} \centi\meter^{-3}$. In the case of our outflows model, a fraction of the mass that should be accreted, hereafter $f_{\mathrm{acc,outflows}}$, is rather ejected in a bipolar cone of opening angle $\theta_{\mathrm{outflows}}$ at a fraction $f_{\mathrm{esc,outflows}}$ of the escape velocity. The direction of this cone is then given by the angular momentum direction of the sinks. We refer the reader to the appendix A.1 of \cite{2022A&A...663A...6V} for the technical details about the implementation of the outflows methods in \texttt{RAMSES}.  In this work, we investigated a single configuration for the outflows, i.e. $f_{\mathrm{acc,outflows}}=1/3$, $f_{\mathrm{esc,outflows}}=1/3$ and $\theta_{\mathrm{outflows}}=20 \degree$. We recall that this parameter choice is based on observational evidences as explained, again, in the appendix A.1 of  \cite{2022A&A...663A...6V}. As they argue, both theoretical \citep[e.g.,][]{1982MNRAS.199..883B,1986ApJ...301..571P} and observational \citep[][]{1995AJ....109.1846H,2007A&A...468L..29C} studies indicate that $f_{\mathrm{acc,outflows}}$ is between 0.1 and 0.4 while $f_{\mathrm{esc,outflows}}$ should range between 0.25 and 0.5. In addition, and as explained by \cite{2021MNRAS.502.3646G}, the effect of outflows (their lever arm) is actually controlled by the product of these two parameters and a value of 0.3 for both parameters puts the product in the middle of the range constrained by observations \citep{2011ApJ...740..107C}.

\section{Results}

\label{sec:results}
\begin{figure*}
  \centering
 \includegraphics[width=
          0.7\textwidth]{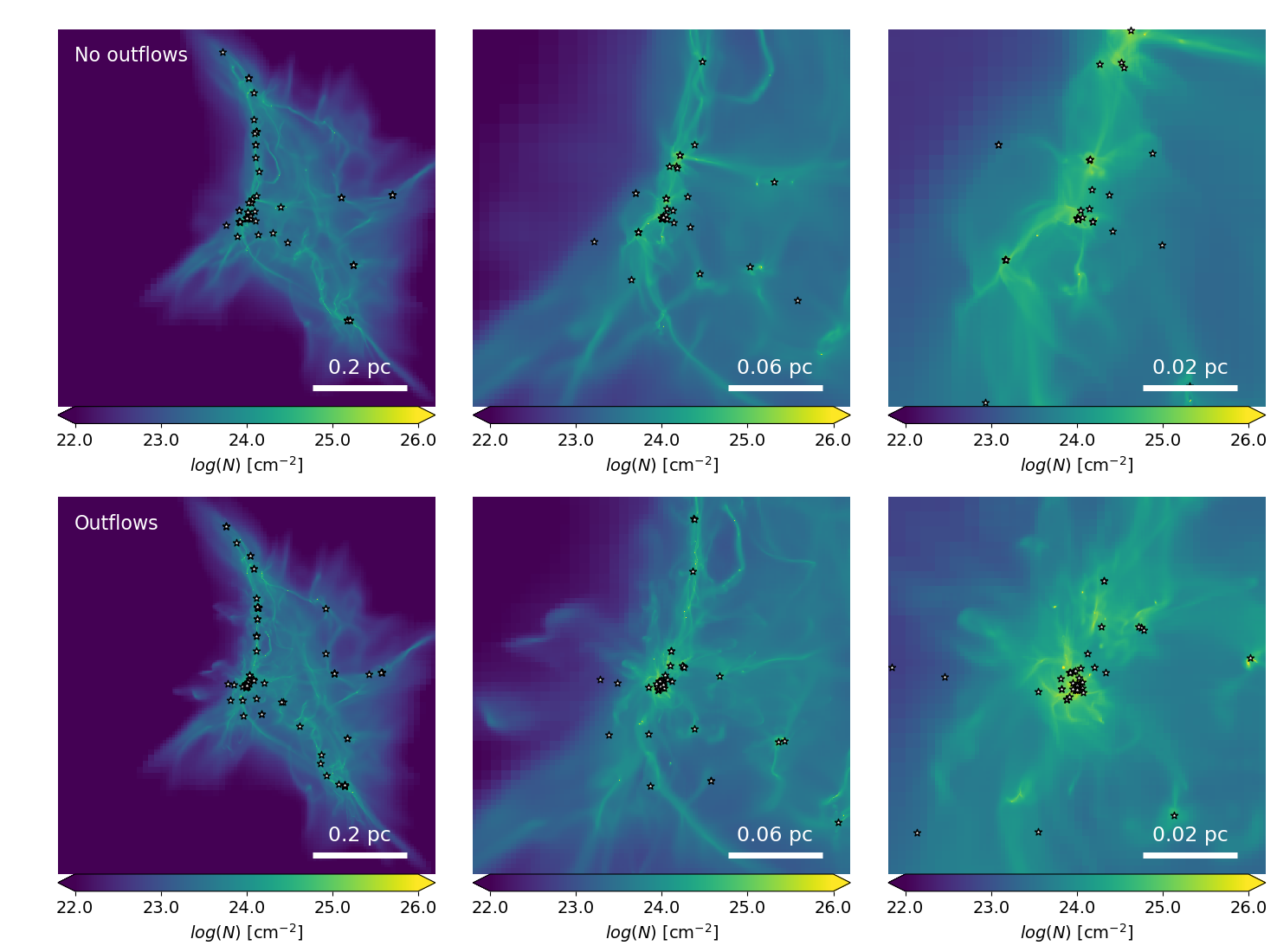}
  \includegraphics[width=
         0.7\textwidth]{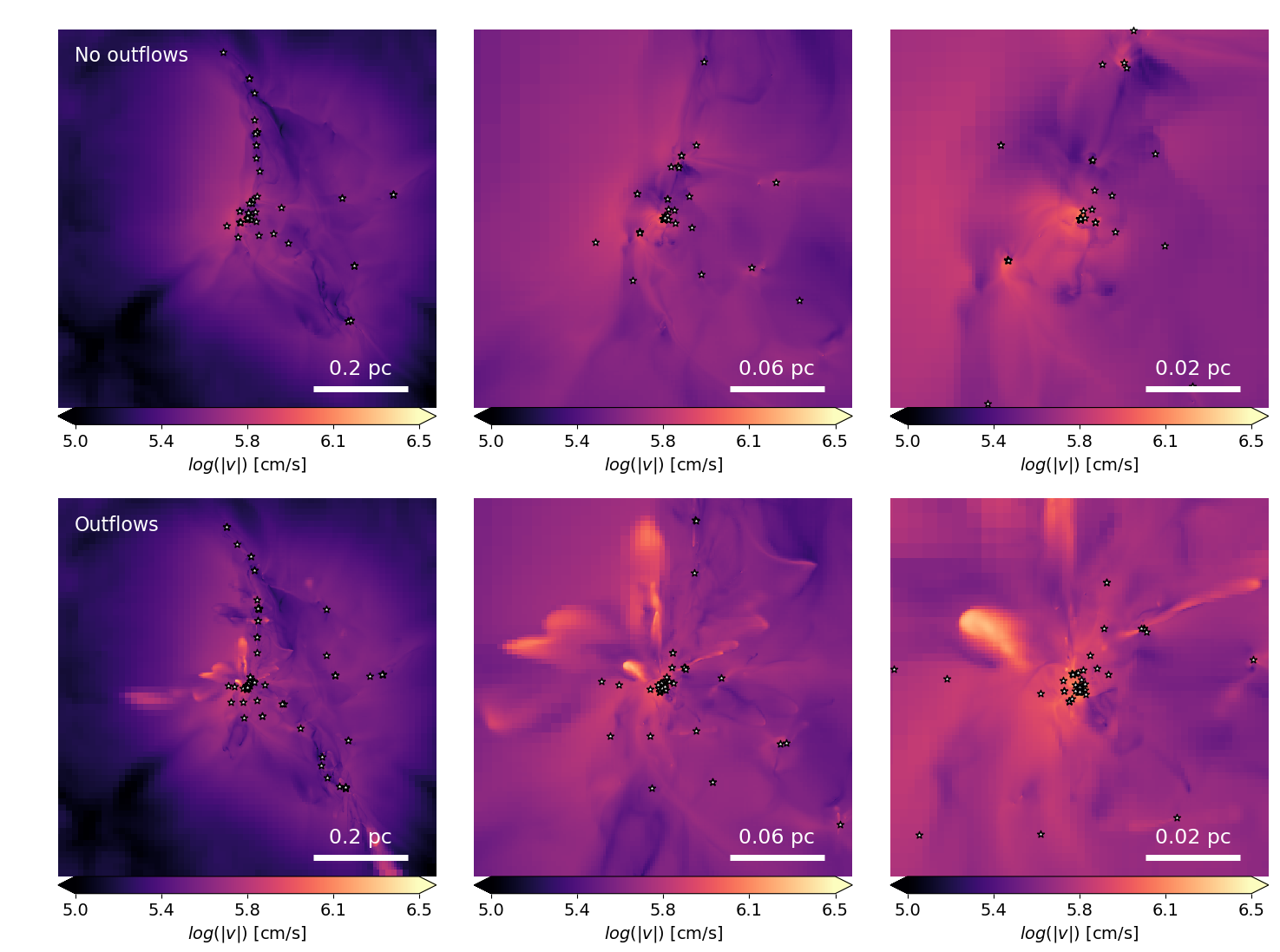}         
\caption{Top 6 maps : \label{fig:column_rho} $(x-y)$ column density maps of the two models integrated along the $z$ direction at SFE=0.1. (Top): without protostellar outflows. (Bottom): with protostellar outflows. From left to right: we zoom toward sink 1 i.e., near the center of the collapse. Outflows have a visible impact on the column density structures are scales below $0.1~$pc. More stars are clearly forming in the presence of outflows. Bottom 6 maps: same but for the mass weighted norm of the velocity. Outflows are significantly modifying the velocity field at scales smaller than $0.1~$pc. }
\end{figure*}
 Let us introduce our two models: \nmhds and \nmhdj. Both are computed according to the methods described in Appendix~\ref{sec:methods}. For \nmhdj, we included the protostellar outflows as described in Appendix~\ref{sec:outflows_methods}, while \nmhds is computed without outflows. We display the evolution of our models in terms of star formation efficiency (SFE), which is the ratio of the mass of formed sink over the initial clump mass. Our models are presented up to SFE=0.1, which corresponds to 100 $M_{\odot}$ that have been accreted within sink particles. As a supplementary material,  we show the evolution of the SFE against time for both models in Fig.~\ref{fig:SFE}. We clearly see that outflows are slowing down star formation as they provide an additional source of mechanical support against the collapse increasing the time reach a given SFE by about $\sim 15-20~\%$.

\subsection{Impact on the large scale structures}

\begin{figure}[h!]
  \centering
 \includegraphics[width=
          0.35\textwidth]{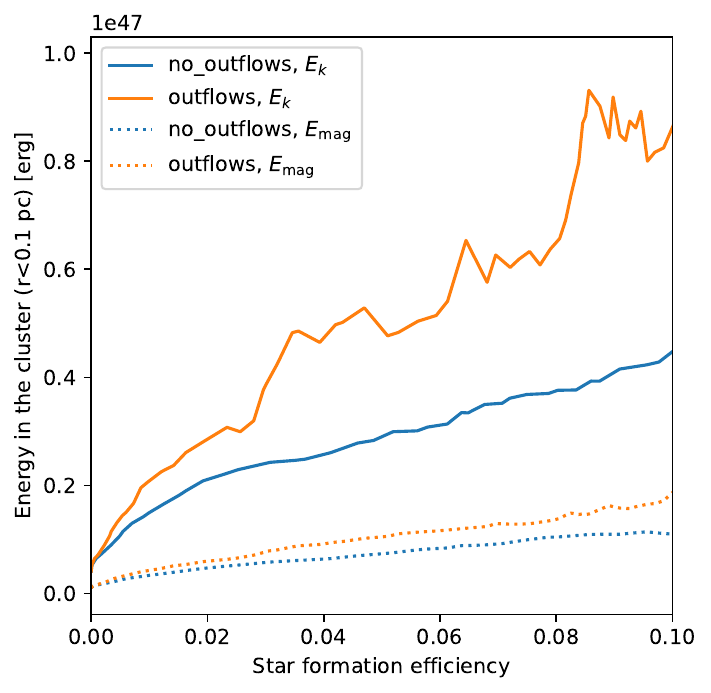}
          \includegraphics[width=0.35
          \textwidth]{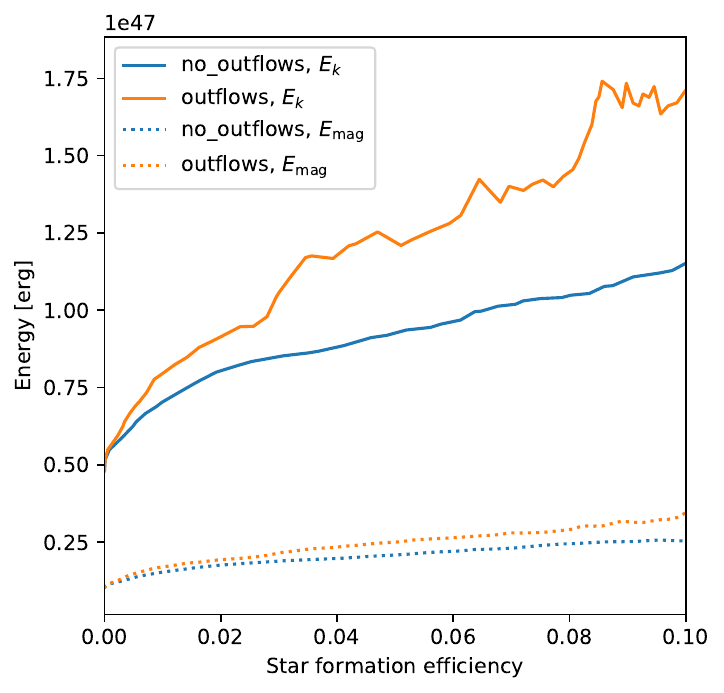}
        \includegraphics[width=0.35
          \textwidth]{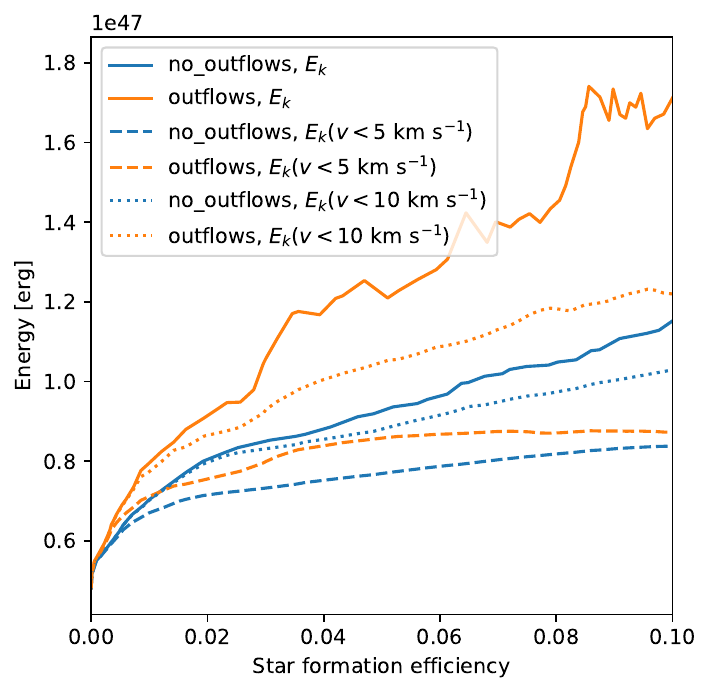}         
          \caption{\label{fig:EK} Top and middle plots: Kinetic (plain lines) and magnetic (dotted lines)  energies as a function of SFE for the two models in a sphere of 0.1 pc surrounding the main cluster (top) and in the full clump (middle). As can be seen, outflows increase the overall support of the cloud as both the kinetic and magnetic energy are larger in their presence. The bottom plots shows the kinetic energy in the cluster with various velocity thresholds. We see that outflows essentially add kinetic energy in a fast $v> 10~\kilo\meter~\second^{-1}$ component.}
\end{figure}

\begin{figure*}
  \centering
 \includegraphics[width=
          0.7\textwidth]{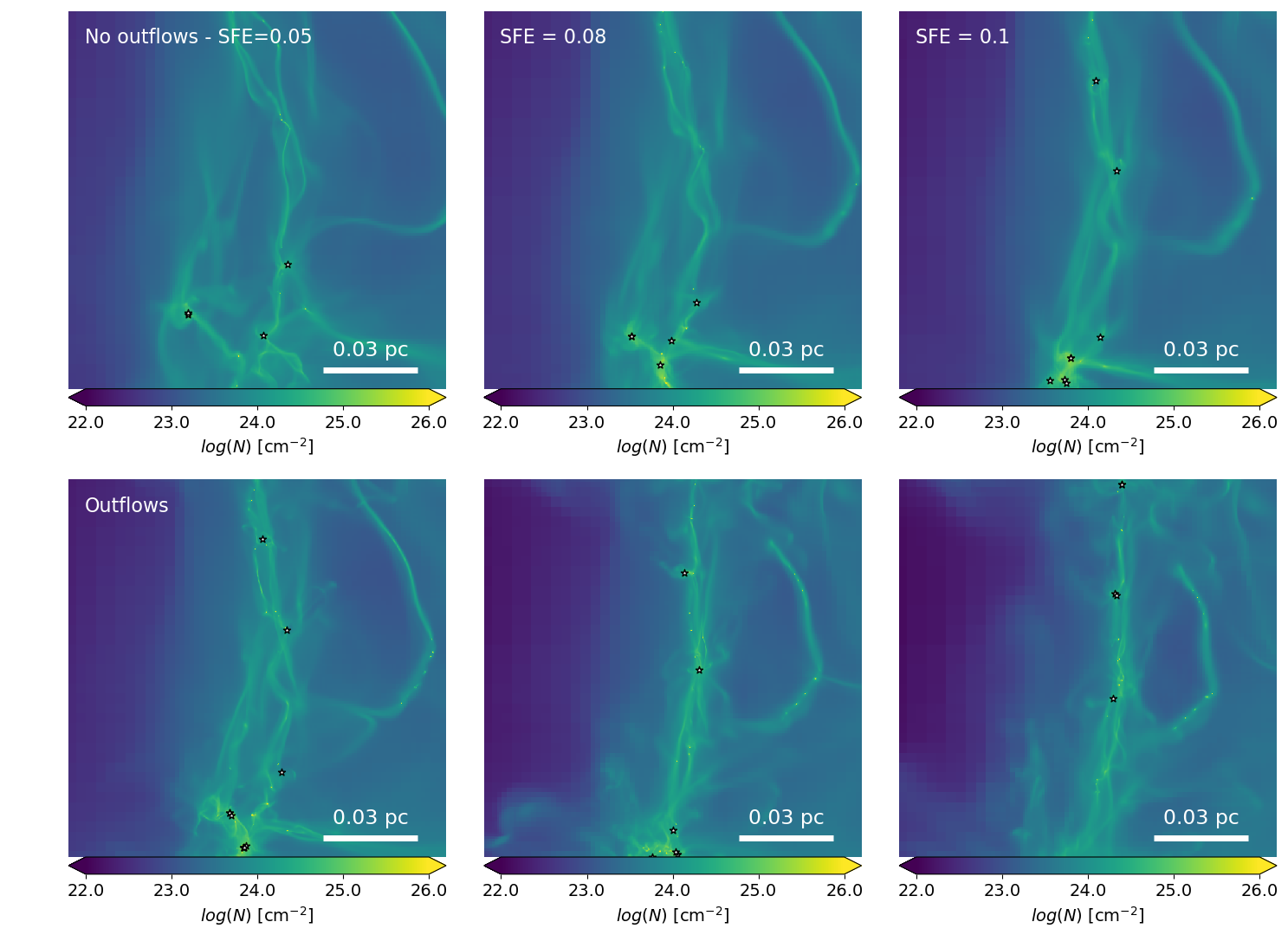}
 \includegraphics[width=
          0.7\textwidth]{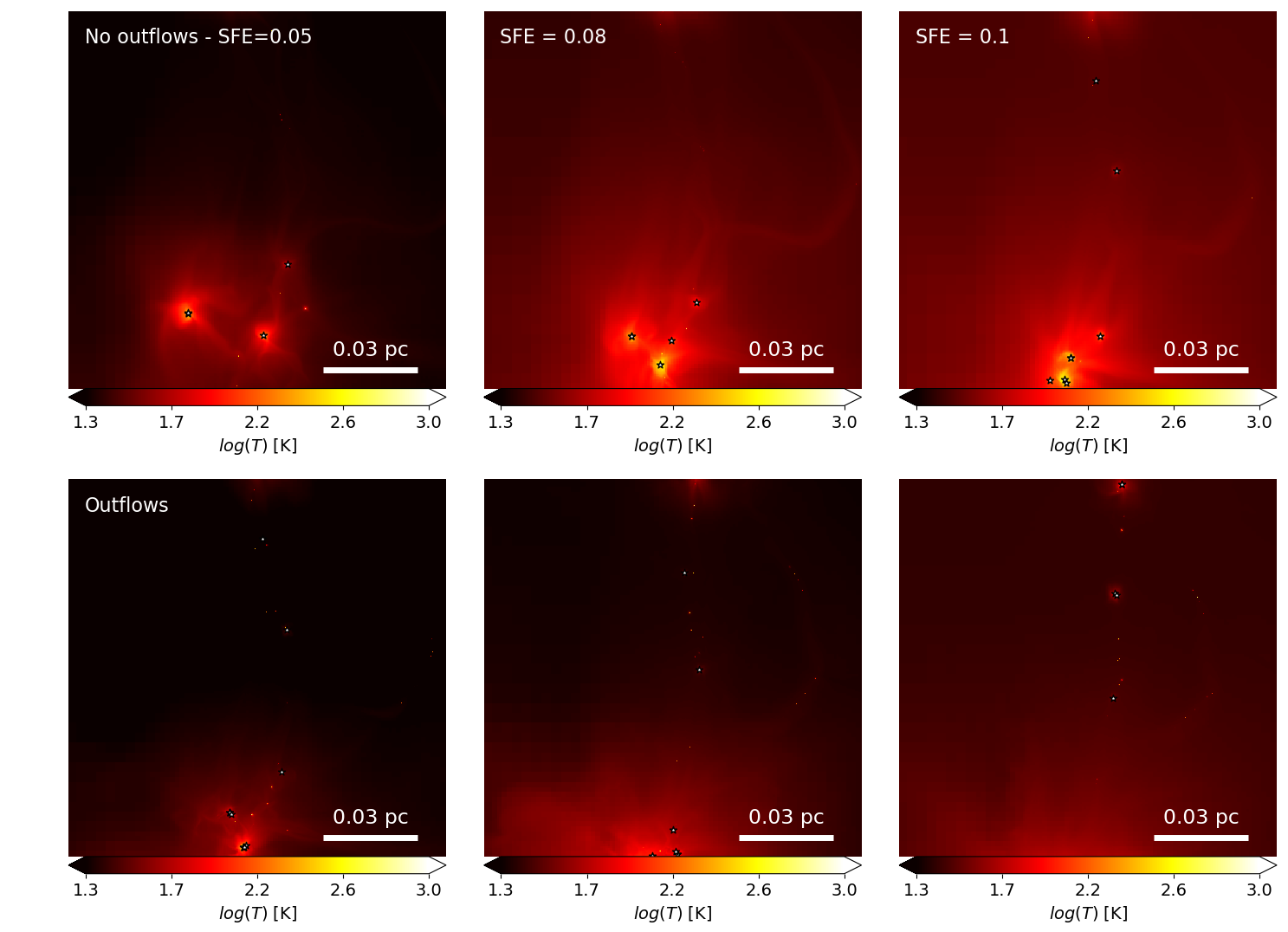}
\caption{\label{fig:filament} Close-up look at a filament located above the main star cluster at three SFEs (0.05,0.08 and 0.1) for the two models. The top panels display the column density and the bottom panels show the temperature integrated along the line of sight.}
\end{figure*}
\begin{figure}
  \centering
 \includegraphics[width=
          0.4\textwidth]{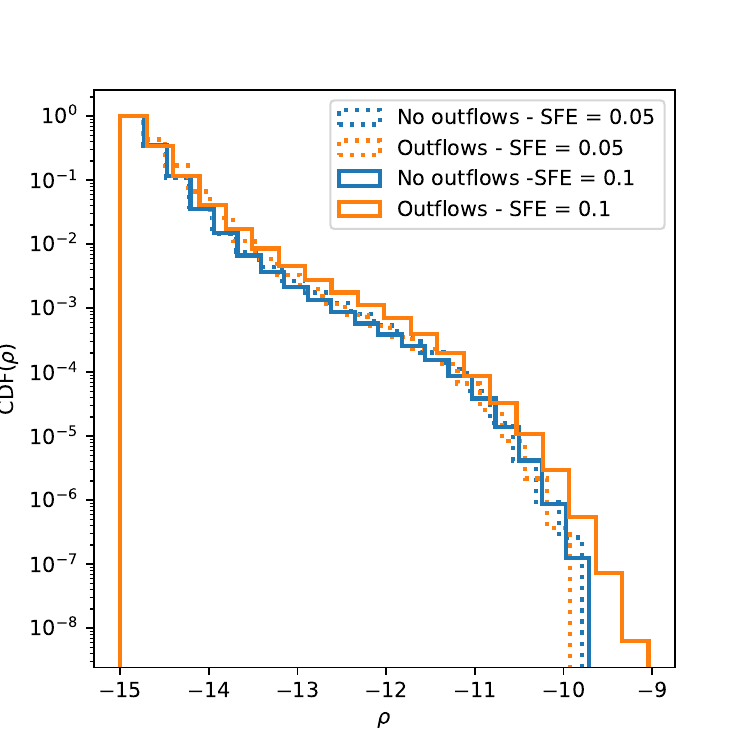}
\caption{\label{fig:CDF} Inverse cumulative distribution function (CDF) of $\rho$ for the dense gas (for $r> 10^{-15} \gram \centi\meter^{-3}$) outside the main star cluster i.e., for r $>0.1$~pc for both models at SFE=0.05 (dotted lines) and SFE=0.1 (plain lines). We clearly see, that outflows promote fragmentation between SFE=0.05 and SFE=0.1 }
\end{figure}

In this section, we present a global description of the \nmhdjs model and compare it with run \nmhd.  In Fig.~\ref{fig:column_rho}, we show the column density and the mass averaged velocity norm integrated along the $z$-direction, respectively, for \nmhds (top) and \nmhdjs (bottom). The images are displayed at SFE=0.1 and centered around sink 1, located near the collapse center. From left to right, we show these maps for decreasing scales.

For both runs, a network of star-forming filaments is clearly noticeable. This is an expected consequence of gravo-turbulent motions.  As already reported in \cite{Lebreuilly2021}, a main star cluster, observed for both models, forms at the vicinity of sink 1. Noticeably, the density maps of the two models are very similar, particularly at large scales ($>0.1$~pc). Similar behavior was also noted in the previous studies of \cite{2011ApJ...740...74K}, \cite{2021MNRAS.502.3646G} and \cite{2022A&A...663A...6V}. We point out that significant differences appear when looking at rather small scales ($<0.1~$pc), also in line with the aforementioned studies. These differences between the two models are more striking near the main star cluster. We clearly see that more stars are able to form there in the case of \nmhdj, this effect is described more in details in Sect.~\ref{sec:IMF}. Now focusing on the velocity maps, we notice that they are also affected by outflows. In particular, we see individual outflows quite distinctively. By the end of the simulation, the affected region reaches a scale of $\sim 0.1~$pc. This is consistent with a propagation at a few $\sim 10~\kilo\meter~\second^{-1}$. This is also in agreement with molecular outflows observations \citep[see e.g.,][]{2009A&A...499..175M,2018MNRAS.476..771C} as well as previous numerical works \citep{2014ApJ...790..128F,2018MNRAS.473.4220L,2021MNRAS.502.3646G,2022A&A...663A...6V}. 

In Fig.~\ref{fig:EK} we show the evolution of the total kinetic (plain lines) and magnetic (dotted lines) energies for the two models against the SFE in a radius of 0.1 pc around sink 1, that we call the main cluster (top), in the full clump (middle) and for various velocity threshold in the cluster (bottom). For both models, the kinetic energy in the main cluster increases over SFE and lies between a  $\sim \times 10^{46}$ and $\sim  10^{47}$~erg (while the total kinetic energy ranges between $\sim 5 \times 10^{46}$ and $\sim  1.8 \times 10^{47}$~erg). As was also reported by previous studies \citep[see][and the other numerical works cited above]{2007ApJ...662..395N,2009ApJ...695.1376C}, outflows are clearly providing additional support for the cloud, operating mostly at small scales. Indeed, a significant amount of kinetic energy is added to the clump by the outflows. Near the end of the calculation, the kinetic energy around the main cluster is about two to three (depending on time and scale) times larger when outflows are included. In total, outflows bring about $\sim 50-70 \%$ more kinetic energy to the full clump. Given that about $40 M_{\odot}$ have been ejected in total by the outflows, the additional kinetic energy of $\sim 5 \times 10^{46}$~erg brought by the outflows to the full clump is consistent with a bulk velocity of a few $ 10~\kilo\meter~\second^{-1}$. Noteworthy the extra kinetic energy brought by outflows is mostly under the form of an high  $ > 10~\kilo\meter~\second^{-1}$ velocity component rather than under the form of turbulence. This can be seen, in the bottom panel of Fig.~\ref{fig:EK}, were we show the kinetic energy of the cluster with various velocity thresholds. The kinetic energies for $v< 5~\kilo\meter~\second^{-1}$ and $v< 10~\kilo\meter~\second^{-1}$ are indeed much closer to each other in the two models. Quite interestingly, the magnetic energy of  \nmhdjs is also increased with respect to \nmhd, outflows are not only providing additional kinetic energy but are also able to increase the magnetic support.

Outflows clearly enhance the fragmentation at the filament scale. This can be seen in Fig~\ref{fig:column_rho} (and Fig~\ref{fig:column_rho2} of Appendix~\ref{appendixCL} which displays the same information but for SFE=0.05). Quite clearly, the \nmhdjs clump is more fragmented and disturbed than \nmhds at both SFE=0.05 and 0.1. This effect is particularly visible at scales below 0.05 pc (right panels) near the main star cluster or in filaments at SFE=0.05. 

Fig\ref{fig:CDF} demonstrates that outflows indeed support fragmentation in the filaments. With this figure, we show the inverse cumulative distribution function (CDF) of density in the dense gas at SFE = 0.05 and SFE=0.1 for the two models. We have excluded the cluster from the CDF, as star formation occurs mostly in the filaments. Here, we clearly see that a significantly larger fraction of the gas is dense (filament, cores and disks) when the outflows are accounted for. In Fig~\ref{fig:filament} we also see that effect quite clearly. In these maps, we show the evolution of the column density for one filament, located right above the main cluster, for three different SFE (0.05, 0.08 and 0.1, from left to right) without (top) and with (bottom) outflows. We clearly see that filaments are more fragmented in \nmhdjs that \nmhd.

In the lower panels of Fig.~\ref{fig:filament}, we show the temperature of the filaments (integrated along the line of sight) at these three evolutionary stages with and without outflows. Quite distinctively, not only the background temperature but more importantly the filament temperature is lowered by the presence of outflows. Without outflows, the filament temperature is increased by up to $\sim 40-50 \%$ at SFE=0.08. As we show in Sect.\ref{sec:disks}, this is a consequence of the lowering of the stellar feedback as result from the reduction of the accretion rates by outflows. Reduced stellar feedback have been shown to favor fragmentation by \cite{2020ApJ...904..194H} and \cite{Hennebelleetal2022}. In fact, in the top panels of Fig~\ref{fig:filament} we even see a thickening of the arc-shape filament (as it gets hotter through the effect of stellar feedback) and a suppression of fragmentation over time whereas the ones of the bottom panels (i.e., with outflows) remain thinner and more fragmented. We point out that the added kinetic energy by the outflows might also be playing a role in that increased fragmentation. Indeed, an enhanced fragmentation was observed by \cite{2010ApJ...709...27W} with isothermal calculations, as such the excess of fragmentation could only come from the modification of the cloud dynamics in their case.

\subsection{Impact on the disk population}
\label{sec:disks}
\begin{figure*}
  \centering
 \begin{subfigure}{ 0.4\textwidth}
  \centering
 \includegraphics[width=
          \textwidth]{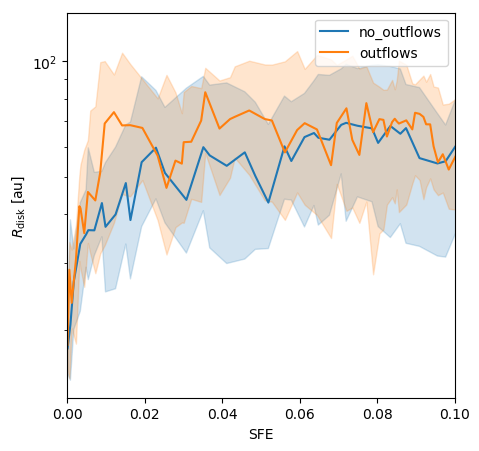}
      \caption{Disk radius}       
     \end{subfigure}
      \begin{subfigure}{ 0.4\textwidth}
  \centering
 \includegraphics[width=
          \textwidth]{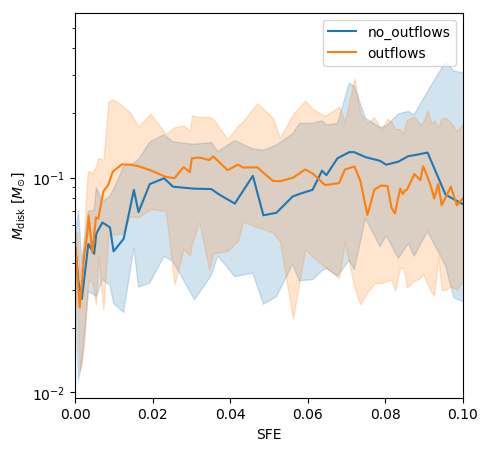}
      \caption{Disk mass }       
     \end{subfigure}
     \\
      \begin{subfigure}{ 0.4\textwidth}
  \centering
 \includegraphics[width=
          \textwidth]{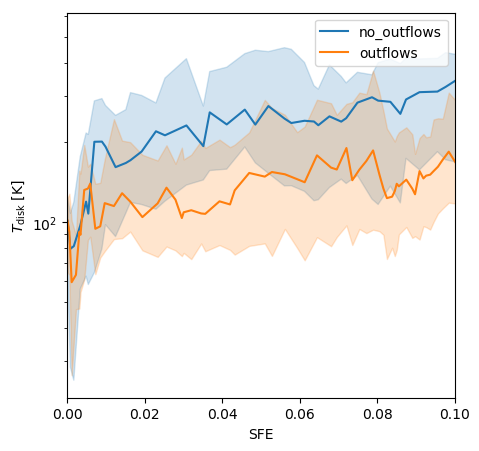}
      \caption{Disk temperature }       
     \end{subfigure}
      \begin{subfigure}{ 0.4\textwidth}
  \centering
 \includegraphics[width=
          \textwidth]{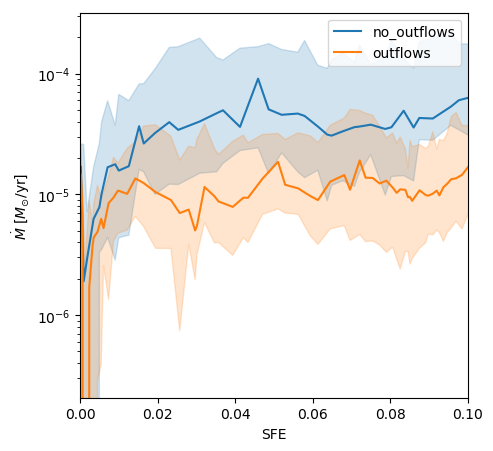}
      \caption{Accretion rate onto the primary }       
     \end{subfigure}
     \caption{\label{fig:pop}Evolution of the median disk  radius, their mass, their temperature and the mass accretion rate onto the primary as a function of the SFE for the \nmhds (blue) and \nmhdjs (orange) models. The transparent colored regions represent the area between the first and third quartile of the distribution. We clearly see that outflows do not impact significantly the disk size and mass but largely influence their temperature through a reduced accretion rate (and luminosity).}
\end{figure*}

\begin{figure*}
  \centering
 \includegraphics[width=
          \textwidth]{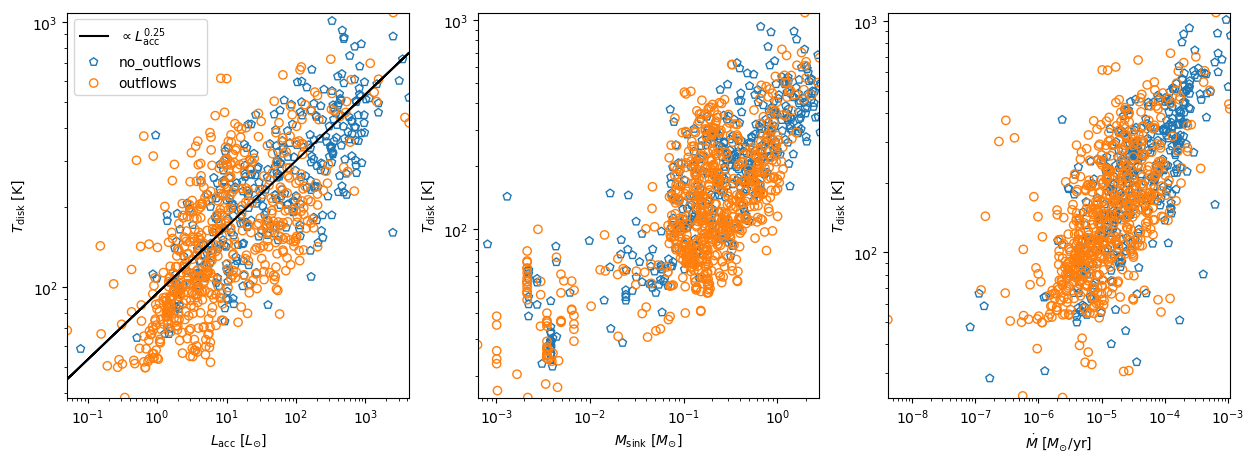}       \caption{\label{fig:acc} Disk temperature vs their accretion luminosity (left), the mass of the sink (middle), the accretion rate onto the sink (right) for both models. All disks are displayed every kyr. The black line denotes the correlation $T_{\mathrm{disk}}\propto L_{\mathrm{acc}}^{0.25}$. The clear correlation confirms that our disks are passively heated by the stars through the accretion luminosity.  }
\end{figure*}

\begin{figure}
  \centering
 \includegraphics[width=
          0.4\textwidth]{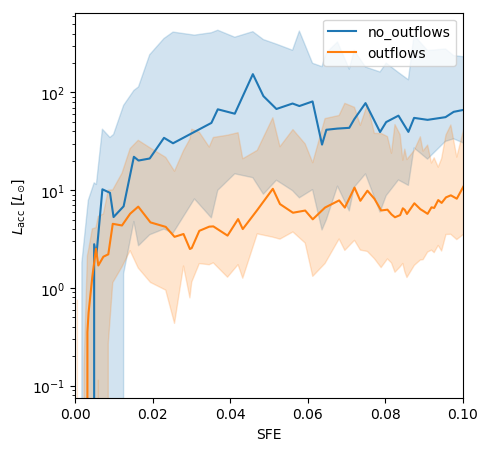}       \caption{\label{fig:acclumsfe} Same as Fig.~\ref{fig:pop} but for the accretion luminosity. We clearly see that this quantity is decreased by about an order of magnitude as a consequence of the decrease accretion rate, decreased stellar mass.}
\end{figure}

We now focus on the impact of the protostellar outflows on the disk populations. We refer the reader to \cite{Lebreuilly2021}, for our method to extract the disk internal properties. However, contrary to \cite{Lebreuilly2021}, we select the disk material using solely two of the \cite{2012A&A...543A.128J} criterion:
\begin{itemize}
    \item $n>n_{\mathrm{disk}}\equiv 10^{9}~\centi\meter^{-3}$, $n$ being the gas number density; 
    \item $v_{\phi}>2 v_{r}$,  $v_{\phi}>2 v_{z}$, $v_{r}$, $v_{z}$ and $v_{\phi}$ being the radial, vertical and azimuthal velocities components in the frame of the analyzed sink.
    \end{itemize}
We do not include any condition regarding the gas pressure as this could arbitrarily remove some of the hot component of the inner disk.  The choice of $n_{\mathrm{disk}}$, although typically used in previous works, remains somewhat arbitrary. As such we discuss it in Appendix~\ref{appendixcrit}.

Once the disk cells are extracted, we average of any given quantity Q in the disk (e.g. the disk Temperature) as follows 
\begin{equation}
    \left<Q\right>= \sum_{j\in\mathrm{disk}} Q_j \Delta x_j / \sum_{j\in\mathrm{disk}} \Delta x_j,
\end{equation}
were $\Delta x_j$ is the cell size. 

We show the evolution of the disk radius (top-left), mass (top-right), temperature (bottom-left) and the accretion rate onto the star (bottom-right) as a function of the SFE for both models (\nmhds in blue and \nmhdjs in orange) in Fig.~\ref{fig:pop}. The lines represent the distribution median and the transparent regions represent the area between its first and third quartile. As such these figures allow to visualize the full disk population as a function of SFE. Some typical protostellar disks are displayed in Appendix~\ref{appendixC} for both runs as a supplementary material.

We first focus on the disk size and masses. Quite clearly, the outflows do not have a significant influence on the two quantities. In term of size, the disks are typically quite compact, with a median radii between $\sim 30$ and $\sim 60$~au depending on the time. The radii evolution is very similar for both models. They are in agreement with the previous models of \cite{Lebreuilly2021} and the observations of disks around Class 0 protostars \citep[e.g.][]{2019A&A...621A..76M,2020ApJ...890..130T,2022ApJ...929...76S}. For both models, the disk masses typically ranges between a few $0.01 M_{\odot}$ and $0.2 M_{\odot}$. Again, outflows do not appear to have a significant impact. The measured disk masses agree quite well with those found in the hydrodynamical calculations of \cite{2018MNRAS.475.5618B} and \cite{2021MNRAS.508.5279E}. We point out that, as noted in \cite{Lebreuilly2021}, our disk masses are typically greater than the minimum solar mass nebula limit \citep[MSMN, ][]{1981PThPS..70...35H} and as such have enough material to form solar-like planetary systems. This is consistent with an early planet formation as a solution for the potential mass-budget problem of T-Tauri disks \citep{2018A&A...618L...3M}. 

Let us now focus on the disk temperature that is most affected by the inclusion of outflows. First of all, it is useful to point out that the accretion luminosity is quite clearly controlling the temperature. In Fig~\ref{fig:acc}, we show the temperature of the disk vs the accretion luminosity (left), as well as vs the sink mass (middle) and the mass accretion rate onto the sink (right). Each marker corresponds to  a disk. The disks are all displayed every kyr from their birth to the end of the simulation. The black line represents the correlation $T_{\mathrm{disk}}\propto L_{\mathrm{acc}}^{0.25}$. Quite clearly a very similar correlation is observed for both models, and the high-temperature part of the plot is not populated for \nmhdjs as outflows prevents too strong accretion to happen. The decrease of temperature for high-$L_{\mathrm{acc}}$ in the outflows case is a combined effect of lower mass and lower mass accretion rates. We indeed clearly see that, for a given mass range, the low temperature quadrant is more populated for the case of \nmhdj. We point out that the aforementioned scaling is expected  in the flux-limited diffusion (FLD) approximation \citep[see][]{Hennebelleetal2022} for a passively irradiated disk. Because the accretion is high at those early stages the disks are generally quite hot, especially in the case of \nmhd. For this model, the disk median temperature quickly reaches high values, around $200-300$~K. Noteworthy, the typical disk temperature is lower by a factor of $\sim 2$ when outflows are included. This is because the accretion rate onto the protostars are significantly reduced when outflows are included. Owing to the relatively weak scaling between the temperature and the accretion luminosity, only an important change in the accretion luminosity would affect the disk temperatures significantly. This is exactly what is happening in our two runs. As can be seen in panel (d), when outflows are included the typical accretion rate decrease by a factor of a few. Consequently, the median accretion luminosity, that scales as $\propto M \dot{M}$, is about one order of magnitude lower when including the protostellar outflows, with a median value of $\sim 50 L_{\odot}$ for \nmhds and $\sim 5 L_{\odot}$ for \nmhdj. This can be seen in Fig~\ref{fig:acclumsfe}, that shows the evolution of the accretion luminosity for the two models as a function of SFE. This is perfectly consistent with the factor of 2 in median disk temperature between the two models according to the scaling previously reported. We point out that the accretion luminosity that we measure are more in line with observations when the outflows are included since YSOs seem to have typical luminosities of a few $L_{\odot}$ \citep{2011A&A...535A..77M,2012ApJ...747...52D,2017ApJ...840...69F}. 
In addition from episodic accretion, outflows could be important to regulate the accretion rate and solve the luminosity problem of YSOs \citep{2011ApJ...736...53O,2012ApJ...747...52D,2022MNRAS.517.4795M,2023MNRAS.518..791E}

\subsection{Impact on the star formation}
\label{sec:IMF}
\begin{figure}
  \centering
 \includegraphics[width=
          0.4\textwidth]{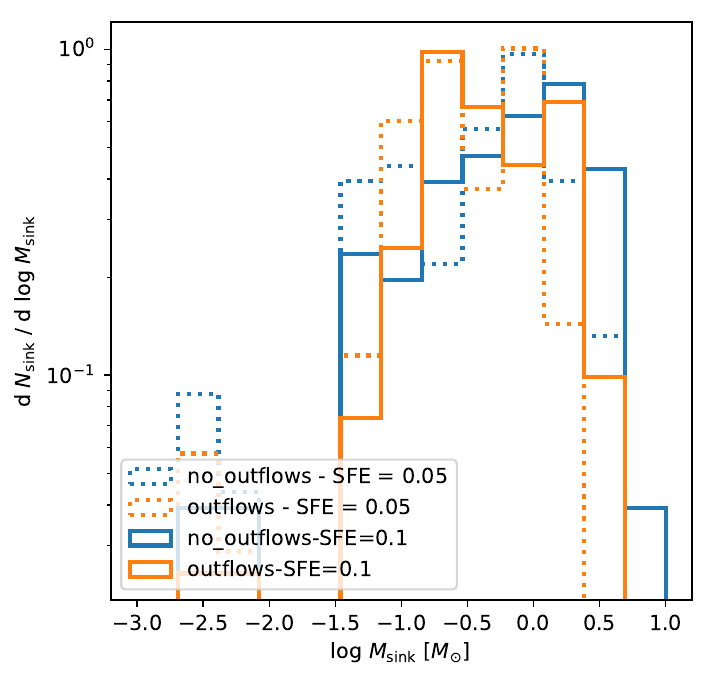}
          \caption{ \label{fig:imf} Stellar mass spectrum of the two models at SFE=0.05 (dotted lines) and SFE=0.1 (plain lines). Outflows do not affect the number of low mass stars, i.e. below few 0.1 $M_\odot$ but affect 
          the number of more massive stars. In particular, they tend to increase the 
          number of stars of mass around 0.3 $M_\odot$ and reduce the most massive star number.  }
\end{figure}

Finally, we describe how star formation proceeds in the clump and how the stellar IMF is affected by the inclusion of protostellar outflows. As previously explained, the two models have been integrated up to a SFE of 0.1 which corresponds to $t_0+38.5$~kyr for \nmhdjs and $t_0+32.5$~kyr for \nmhd, $t_0=78~$kyr being the time of formation of sink 1. Before this time the two models are identical. Protostellar outflows are reducing the time needed to achieve a given SFE by $15-20~\%$. We also clearly see in Fig.~\ref{fig:SFE} that, at around $t=110~$kyr, the SFE is of the order of 0.1 for \nmhds and 0.05 for \nmhdjs, meaning that the star formation rate (SFR) is reduced by a factor $\sim 2$ by the outflows, which is consistent with the accretion rates that we report in this article.  Physically, this is not a surprise as outflows imply that a fraction of the mass above the accretion threshold is now ejected rather than accreted.  As pointed out earlier, in addition from slowing down accretion, the inclusion of outflows also increases fragmentation, this allows the clump to form significantly more stars. Indeed, by SFE=0.1, \nmhdjs has formed 133 sinks while \nmhds has only formed 84. We point out that, for both models, star formation mainly occurs in the dense filaments of the clump. Even the main star cluster originates from filaments, which are later destabilized by the global dynamics of the clump and, in the case of \nmhdjs, by the outflows.

We now show, in Fig.~\ref{fig:imf} the IMF of the two clumps at SFE=0.05 (dotted lines) and SFE=0.1 (plain lines). First of all, we note that outflows do not seem to visibly shift the low mass stars part of the IMF, that is to say the stars of mass below 0.1 $M_\odot$.  Indeed, as we resolve the first Larson core mass which, as shown by \cite{2019ApJ...883..140H} is a good candidate as a mass scale for the IMF, we are able to probe the transition from an isothermal to an adiabatic equation of state. This stops the fragmentation in lower mass objects because at least one first hydrostatic core mass (about 0.03 M$_\odot$) is required at the high densities for the introduction of sink particles. 

Outflows however affect the shape of the high-mass part of the IMF or, as argued by, \cite{2018MNRAS.473.4220L,2021MNRAS.502.3646G,2023MNRAS.518.5190M}, its peak. 
We stress that the differences we observe between the IMFs obtained with and without outflows, are significantly 
lower than what is reported for instance in Fig.~6 of \citet{2021MNRAS.502.3646G}. This is likely because in our simulations
the transition from the isothermal to adiabatic regime is resolved. Without outflows, the median stellar mass is about 1 solar mass and the IMF is quite flat between $\sim 0.2$ and $\sim 2 M_{\odot}$, in agreement with our previous calculations \citep{Hennebelleetal2022}. When outflows are included, the IMF is not top-heavy anymore, and we observe a more distinctive peak around $0.1-0.3~M_{\odot}$. Outflows also appear to play a regulating role for the mass of the most massive stars in cluster $M_{\mathrm{max}}$ and more generally, the mass of the stars above $\sim 0.3 M_{\odot}$. Here,  $M_{\mathrm{max}}\sim 9.4 M_{\odot}$ for \nmhds and it is only $4.1 M_{\odot}$ for \nmhdj. We believe, this could be due to the transition from a regime where thermal/magnetically to a turbulence/kinetically supported regime, as predicted by the model of \cite{Hennebelleetal2022}.

As shown in their appendix B, at a given length $R$, the transition mass between the two regimes scales as 
\begin{equation}
 M_{\mathrm{crit}} \propto \left(c_{\mathrm{s}}^2  R+ \frac{\sigma^2}{3} (R/R_0)^{2\eta} R\right),
\end{equation}
where $\sigma$ is the RMS velocity dispersion at the cloud scale $R_0$ and $c_{\mathrm{s}}$ is the soundspeed. At the transition between the regimes, the terms are roughly equal. As such we have 
\begin{equation}
  c_{\mathrm{s}}^2 \sim  \frac{\sigma^2}{3 } (R/R_0)^{2\eta}, 
\end{equation}
or
\begin{equation}
 R \propto c_{\mathrm{s}}^{1/\eta} \sigma ^{-1/\eta}.
\end{equation}
This yields a critical mass $M_{\mathrm{crit}}$ for the transition between the two regimes which scales as
\begin{equation}
 M_{\mathrm{crit}} \propto c_{\mathrm{s}}^{2+1/\eta} \sigma^{-1/\eta}.
\end{equation}

Assuming a typical scaling $\eta=0.5$ , shows that this critical mass scale quadratically with the temperature, which is a strong dependency. In addition, it scales as the inverse square root of the turbulent velocity dispersion. Both reducing the temperature and increasing the turbulent velocity dispersion would result in a shift toward lower masses of the critical mass.
As shown before, outflows are reducing the clump and filament temperature by up to $40-50\%$ (atb SFE=0.08). Indeed adding them reduces the accretion rate, therefore also affects the stellar feedback by lowering the accretion luminosity and the temperature at filament scales. This reduction, would thus shift the critical mass by a factor $\sim 2$ which is consistent with what we see here. Second, this effect could be helped by the considerable amount of kinetic energy brought by the outflows at small scales. Although strictly speaking the bulk of the kinetic energy that is added by the outflows does not seem to be in the form of turbulence, it still seem to provide a support against the collapse. This is particularly clear in the main star cluster, where outflows are quite visibly modifying the structures. We also note that, as the outflow kinetic energy scales as $\propto M_{\mathrm{sink}}^2$, they are expected to have a stronger influence around more massive objects. It is not clear yet which of the two effects plays the most important role in shaping the IMF, as their both work toward the same direction. Given the scaling of the transition mass demonstrated above, we believe that the temperature effect could be more significant. Future models exploring the impact of the outflow properties, assuming different assumptions for the radiative transfer ($f_{\mathrm{acc}}$) and exploring various cloud configurations while still resolving the disk scales should be dedicated to understanding this effect in more details and to see whether outflows systematically have a strong influence. We also emphasize that our $\sim 1~$au resolution unfortunately comes with a high numerical cost and therefore prevented us so far to produce large statistical samples of sinks or to run a large number of these models.

\section{Conclusion}
\label{sec:conclusions}
In this article, we have presented the first simulations of massive star-forming clumps that resolve the protoplanetary disk scales while including ambipolar diffusion, radiative transfer with stellar feedback following the radiation from internal and accretion luminosity as well as the mechanical momentum input from outflows. We now recall our main findings

\begin{enumerate}
    \item Protostellar outflows have a clear impact on the velocity structures in clumps at scales smaller than $\sim 0.1~$pc and, to a smaller extent, on column density structures. Beyond their propagation scale, their impact is not distinguishable.
 
    \item With or without protostellar outflows, a population of stars and disks is formed in the star-forming clumps. In both cases, disks are born quite compact because of the magnetic braking, in agreement with our previous findings \citep{Lebreuilly2021}. In addition, disks are born massive enough to form planets according to the minimum solar mass nebula criterion. Protostellar outflows do not affect significantly the size and mass of the nascent disks. 

    \item Outflows reduce the typical accretion luminosity in the cloud by about one order of magnitude and could therefore be a potential solution for the well known luminosity problem in YSOs, in addition from episodic accretion \citep{2011ApJ...736...53O,2012ApJ...747...52D,2022MNRAS.517.4795M,2023MNRAS.518..791E}. 
    \item The typical temperature of protoplanetary disks is lower by a factor $\sim 2$ when outflows are included as a consequence of the lower the accretion luminosity.

    \item Protostellar outflows affects the stellar population allowing to form more stars. They also affect the stellar spectrum for masses larger than $\sim 0.3 M_{\odot}$ allowing it to switch from a rather flat (top-heavy) regime to a shape more consistent with the Galactic IMF. Importantly, the mass of the most massive star is reduced by a factor of $\sim 2$ when they are included as they impact the stellar radiative feedback by reducing the accretion onto the protostars and provide additional kinetic support against the collapse. We also find that the low mass star population is relatively unaffected by the presence of outflows. We believe this is due to the fact that this population is, in fact, controlled by the mass of the first hydrostatic core.

\end{enumerate}
    
Ubiquitous around YSOs, protostellar outflows appear to be playing an important regulating role both for star and disk formation by providing an additional form of support as well as by reducing the impact of the accretion luminosity onto the protostars. We strongly encourage future works dedicated to exploring the influence of the unconstrained outflows parameters such as the fraction of ejected material, the outflows launch velocity and their opening angle while still resolving the protoplanetary disks scales.

\appendix
\section{Clump structure at SFE=0.05}
\label{appendixCL}
\begin{figure*}
  \centering
 \includegraphics[width=
          0.7\textwidth]{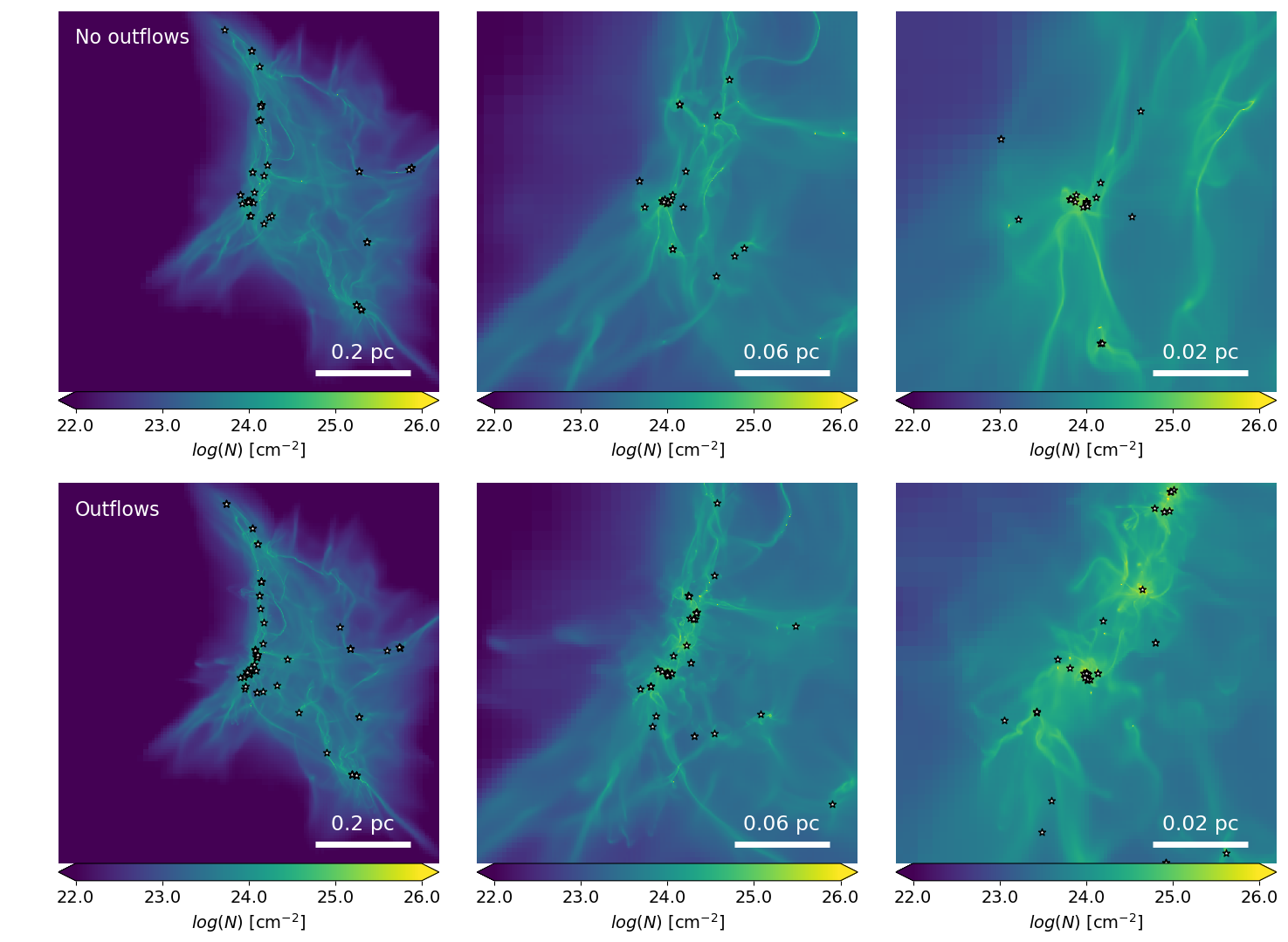}
\caption{\label{fig:column_rho2} Same as Fig.\ref{fig:column_rho} but at a SFE=0.05.}
\end{figure*}

As a complement to Fig~\ref{fig:column_rho}, we show in this section the column density for SFE=0.05 for both models. The impact of outflows is already very visible at that stage, especially at scales below 0.1 pc.

\section{Protoplanetary disk gallery}
\label{appendixC}

In this section, we show the column density and mass weighted velocity integrated along the line of sight of a few typical protoplanetary disks for run \nmhds and \nmhdj. The disks are displayed edge-on at SFE=0.1. The impact of outflows is very clear in the velocity maps but quite invisible in column density maps (as outflows are mostly composed with low density material).

\begin{figure*}
  \centering
 \includegraphics[width=0.7\textwidth]{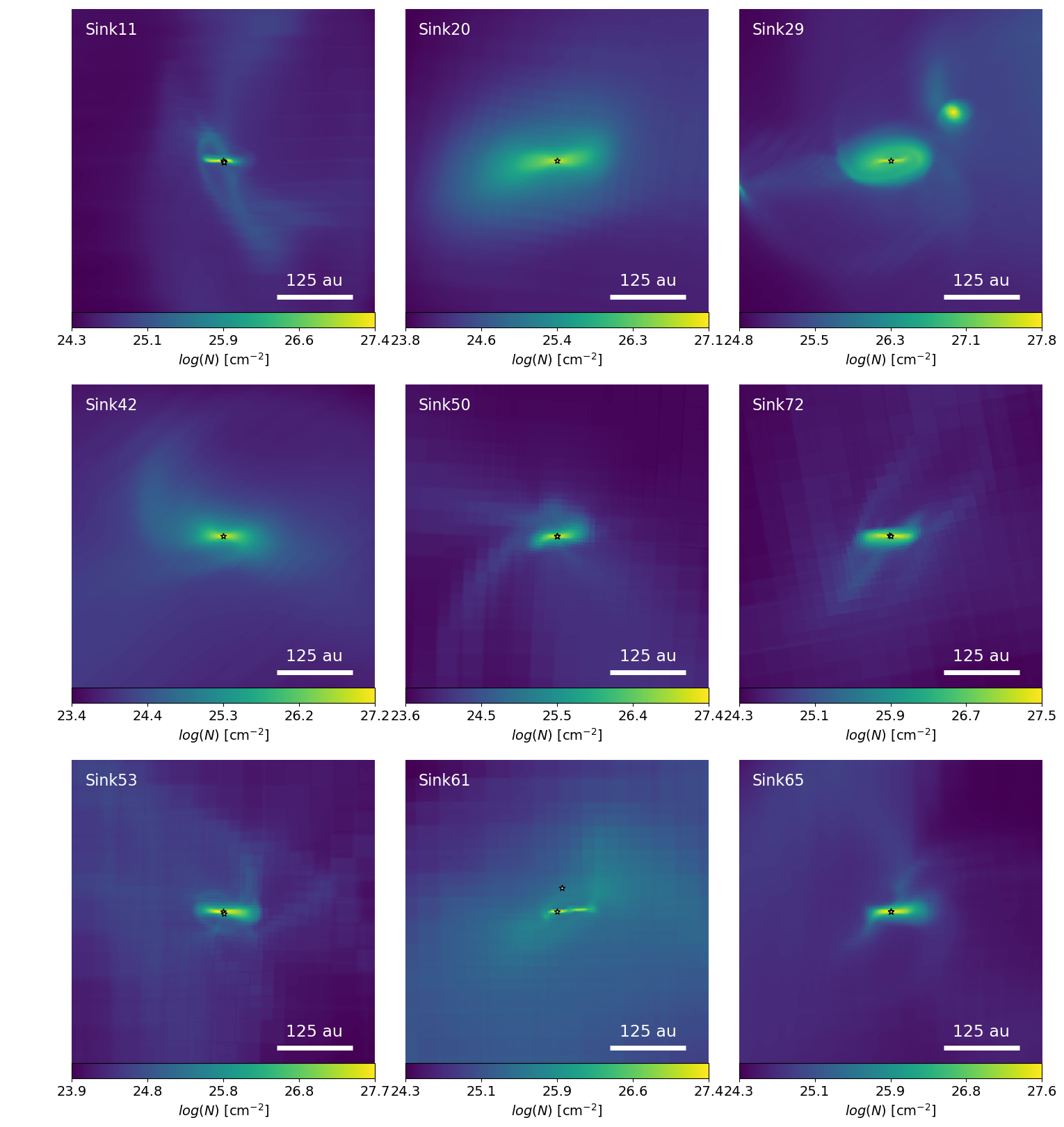}       \caption{\label{fig:disk1} Column density of a few disks seen edge-on for run \nmhds at SFE=0.1.}
\end{figure*}

\begin{figure*}
  \centering
 \includegraphics[width=0.7\textwidth]{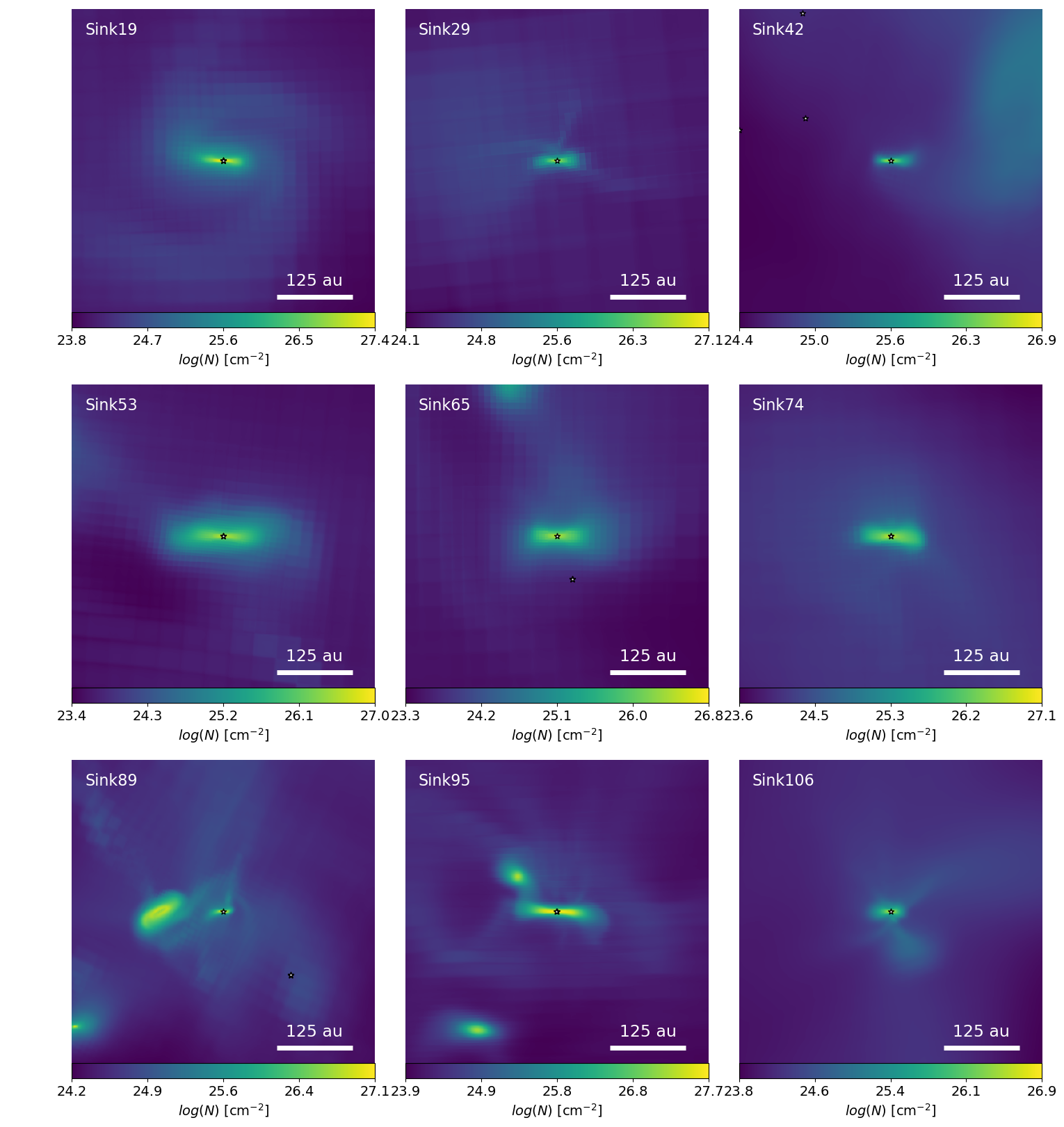}       \caption{\label{fig:disk2} Same as Fig.~\ref{fig:disk2} but for \nmhdj.}
\end{figure*}

\begin{figure*}
  \centering
 \includegraphics[width=0.7\textwidth]{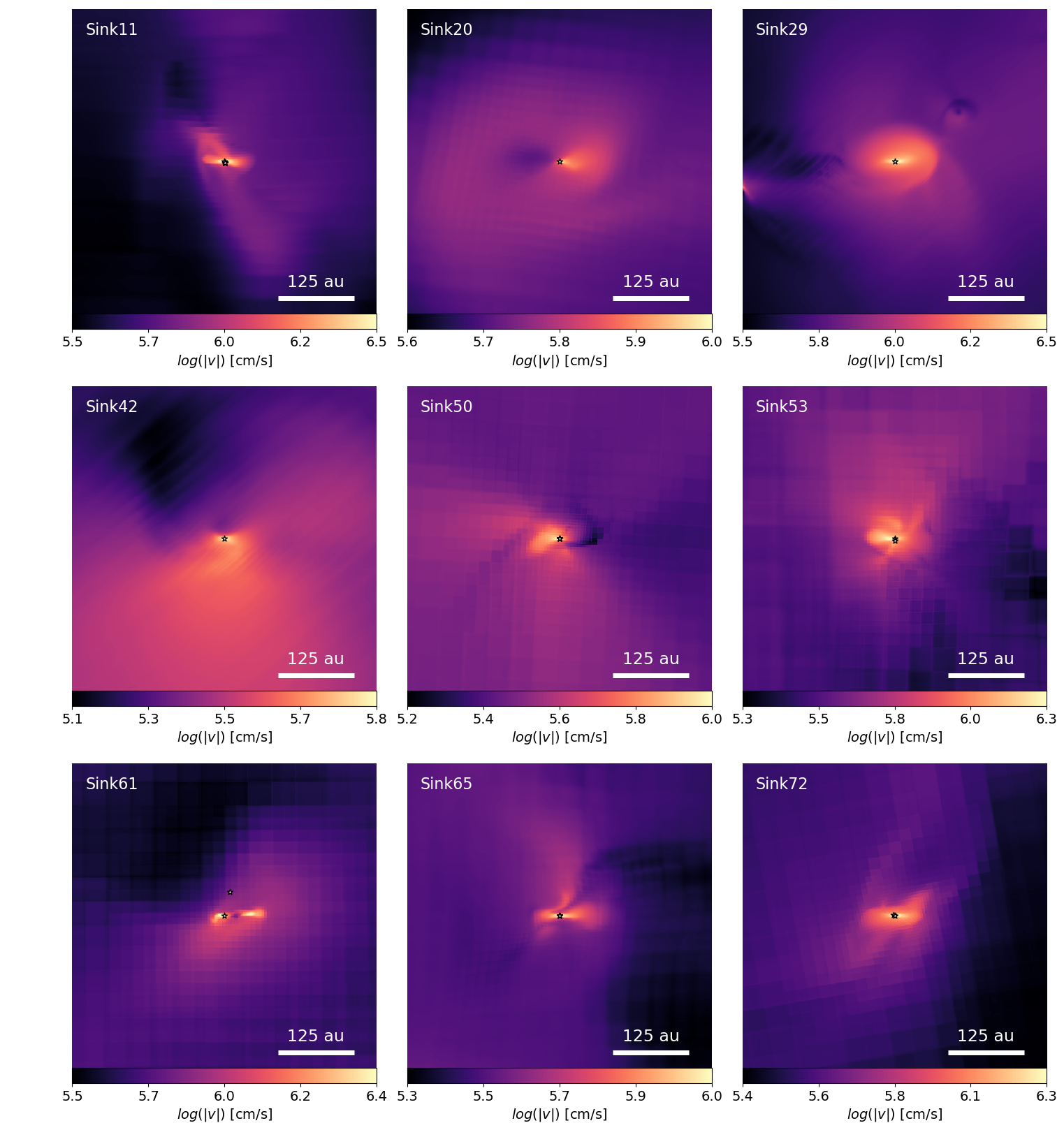}       \caption{\label{fig:disk3} Same as Fig.~\ref{fig:disk1} but for the absolute mass weighted norm of the velocity integrated along the line of sight.}
\end{figure*}

\begin{figure*}
  \centering
 \includegraphics[width=0.7\textwidth]{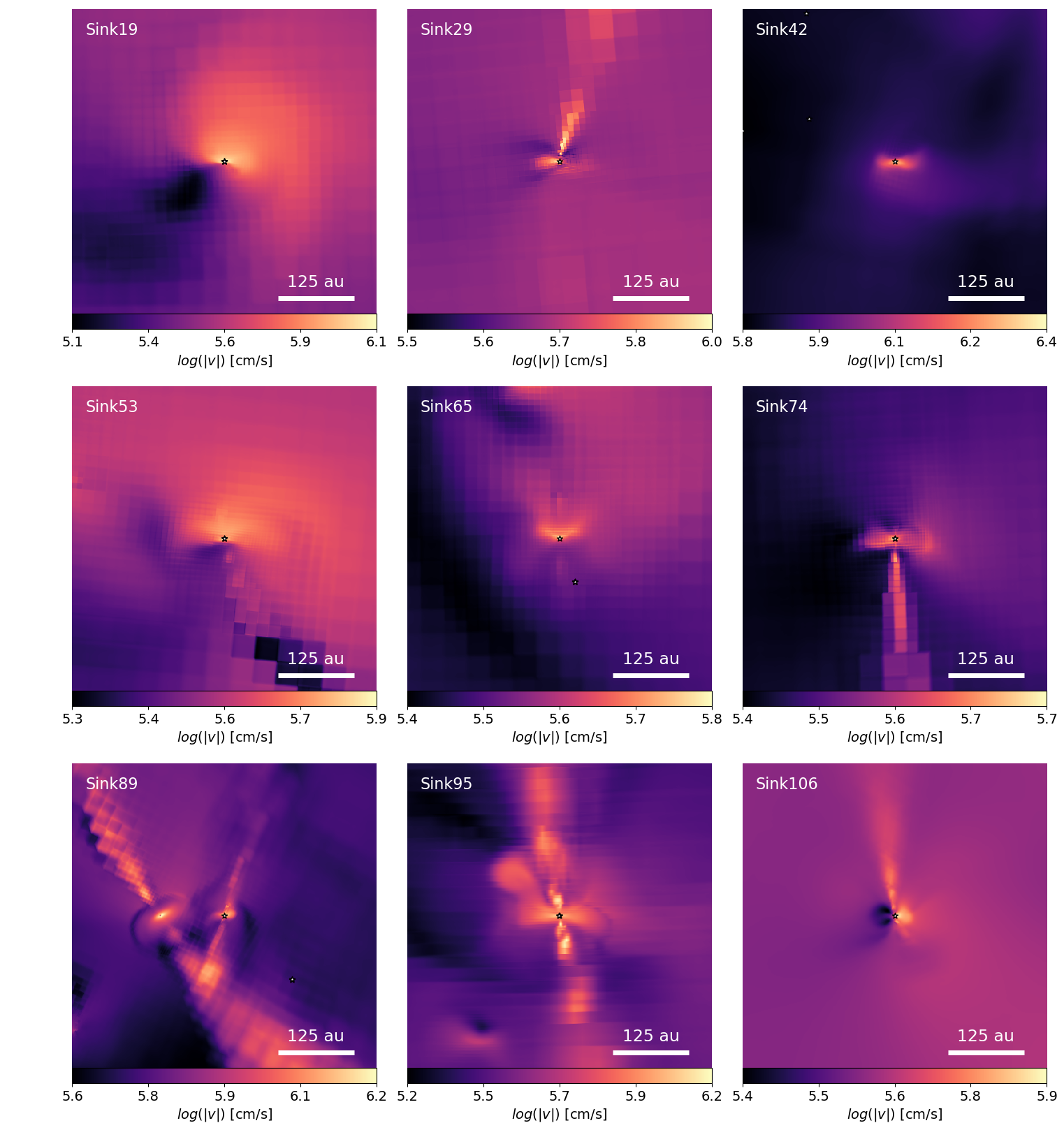}       \caption{\label{fig:disk4} Same as Fig.~\ref{fig:disk2} but for the mass weighted norm of the velocity integrated along the line of sight. The outflows clearly have a visible impact on the velocity field.}
\end{figure*}

\section{Influence of the disk selection density criteria}
\label{appendixcrit}
\begin{figure*}
  \centering
 \begin{subfigure}{ 0.4\textwidth}
  \centering
 \includegraphics[width=
          \textwidth]{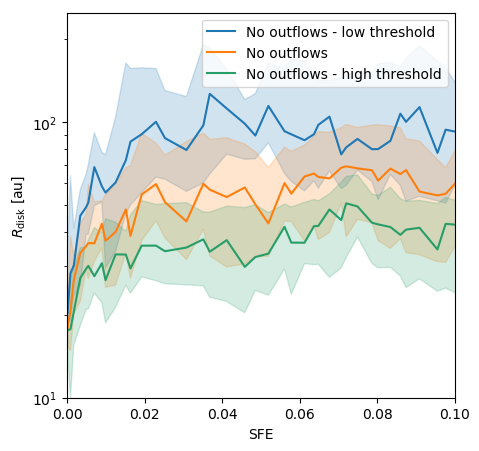}
      \caption{Disk radius: \nmhd.}       
     \end{subfigure}
      \begin{subfigure}{ 0.4\textwidth}
  \centering
 \includegraphics[width=
          \textwidth]{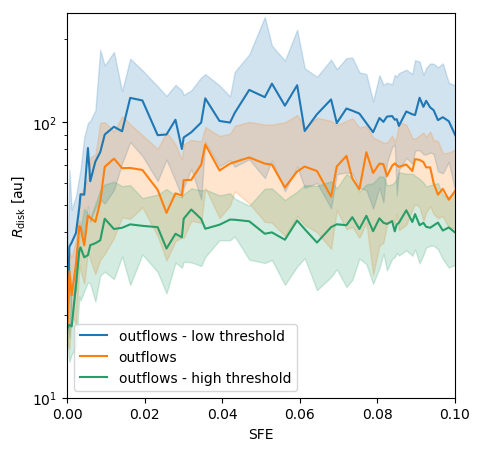}
      \caption{Disk radius: \nmhdj.}       
     \end{subfigure}
     \\
      \begin{subfigure}{ 0.4\textwidth}
  \centering
 \includegraphics[width=
          \textwidth]{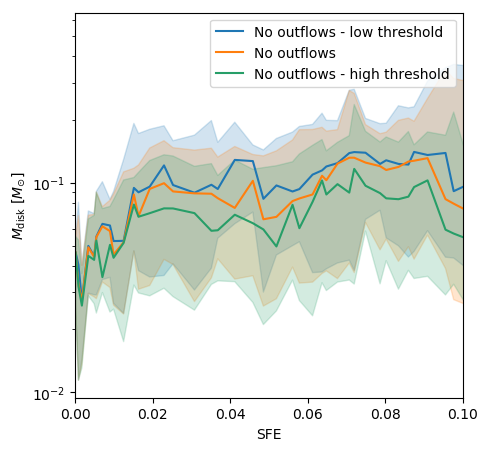}
      \caption{Disk mass: \nmhd. }       
     \end{subfigure}
      \begin{subfigure}{ 0.4\textwidth}
  \centering
 \includegraphics[width=
          \textwidth]{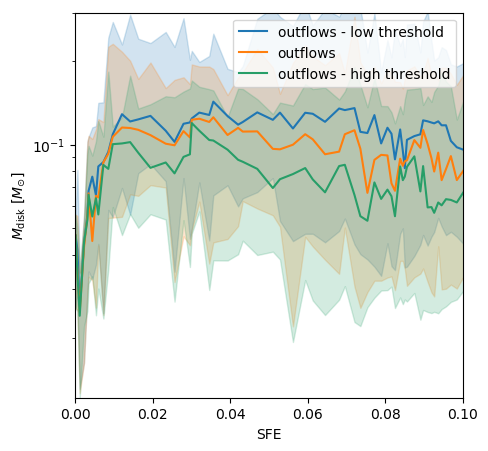}
      \caption{Disk mass: \nmhdj.}       
     \end{subfigure}
     \caption{\label{fig:thre_test} Disk radius (left) and mass (right) for \nmhds (top) and \nmhdjs (bottom) as a function of the SFE for three density threshold for the disk selection.}
\end{figure*}

The density criterion of \cite{2012A&A...543A.128J}, although useful to separate the disk material from the envelope, remains somewhat arbitrary. Therefore, we have varied it by one order of magnitude above and below its reference value of $n_{\mathrm{disk}}=10^{9}\centi\meter^{-3}$ to test its influence on the final estimate of the disk size and mass. We show in Fig.\ref{fig:thre_test}, the evolution of the disk radius (left) and mass (right) as a function of the SFE for the \nmhds (top) and \nmhdjs (bottom) models. As can be seen, the mass spectrum is only weakly affected by the change of density threshold, while the disk size is more significantly impacted. In particular, when going from a threshold of $10^9~\centi\meter^{-3}$ to a threshold of $10^8~\centi\meter^{-3}$ we see that the disk median radius almost shifts toward 100 au. The difference between $10^{9}\centi\meter^{-3}$ and $10^{10}\centi\meter^{-3}$ is however much lower (with a shit from $\sim 50~$au to about $30-40$~au. We not that, by eye, most disks seem to typically have sizes around $\sim 50~$au. As such, $10^{9}~\centi\meter^{-3}$ and $10^{10}~\centi\meter^{-3}$ seem to be a more accurate representation of disks. For continuity with previous studies, we stick to the value $10^{9}~\centi\meter^{-3}$.  We point out that our choice of disk selection criterion does not matter for the comparison between models provided that the same method is used to compare both calculations. This could be more problematic when comparing with observed disks. Synthetic observations of the models are needed to really make a one to one comparison, this dedicated study is currently in preparation.

 \begin{acknowledgements}
We thank the referee for providing a very constructive report that helped us a lot improving our manuscript. This research has received funding from the European Research Council synergy grant ECOGAL (Grant : 855130). We acknowledge PRACE for awarding us access to the JUWELS super-computer.This work was also granted access to HPC resources of CINES and CCRT under the allocation A0130407023 made by GENCI (Grand Equipement National de Calcul Intensif). MG acknowledges the support of the French Agence Nationale de la Recherche (ANR) through the project COSMHIC (ANR-20-CE31- 0009).
\end{acknowledgements}
\bibliographystyle{aa}
\bibliography{ref}

\end{document}